\documentclass{article}

\usepackage[a4paper, total={7in, 9.5in}]{geometry}
\usepackage{graphicx,color}
\usepackage{amsmath,amssymb}

\title{Transverse impact of a Hertzian sphere with an infinite Rayleigh beam}

\author{Saranya Biswas \and Anindya Chatterjee} 

\date{Indian Institute of Technology, Kanpur 208016, India.}

\begin{document}
	
\maketitle


\begin{abstract}
	We study the impact of a Hertzian sphere on an infinite Rayleigh beam, which presents both interesting possibilities for numerical approximation as well as new insights in the multiple-contact-event regime. Here, Galerkin projection using an abstract continuous variable replaces the history-dependent convolution in the impact equations by a system of local-in-time ordinary differential equations. The corresponding approximate solutions are validated against direct numerical solution of the convolution-based formulation and show agreement. Unlike the corresponding Euler-Bernoulli beam impact problem, which is governed by a single nondimensional parameter, the Rayleigh beam impact dynamics is governed by two nondimensional parameters. A parametric study  reveals three distinct impact regimes: one regime with a single contact followed by permanent separation, a second regime with a single contact that is sustained thereafter, and an intervening regime with initial separation followed by multiple contact events with diminishing amplitudes, leading eventually to sustained contact. Short- and long-time asymptotic analyses are used to characterize the boundaries between the permanent separation and the re-contact regimes in different parameter regimes. The three significant new aspects of this paper are the system itself (Rayleigh beam instead of Euler Bernoulli beam), the model order reduction achieved through Galerkin projection in an abstract domain, and the detailed parametric study leading to a simple qualitative picture of possible behaviours.
\end{abstract}


\section{Introduction} \label{intro}

The impact of a compact body against a large flexible structure, such as a plate or beam, is a classical problem in structural dynamics \cite{goldsmith1960impact,jones2011structural,LIN1991293,stronge2000impact}. Finite structures are commonly modelled using lumped-parameter or modal representations \cite{LIN1991293}. However, even for a single Hertzian ball striking an undamped Euler-Bernoulli beam, well over sixty modes may be required for the numerically computed response to converge \cite{bhattacharjee2017interplay}. When the impacting body is small and the contact duration is short, the response near the impact point is instead governed primarily by wave propagation and becomes insensitive to the overall size, shape, and boundary conditions of the structure \cite{olsson2000mass}. The structure can then be approximated as infinite \cite{zener1941intrinsic,schwieger1965simple,CARESTA2023103194}. This idealization eliminates boundary reflections and isolates the interaction between the contact nonlinearity and the structural dynamics, making the problem both more tractable and physically informative.

The infinite-structure idealization also has an important physical consequence. Waves radiated from the contact region do not return, so the impactor can lose energy to the structure even when both are perfectly elastic and undamped \cite{zener1941intrinsic,hunter1957energy,reed1985energy}. The coefficient of restitution is therefore determined by the coupled contact and outward-propagating wave dynamics rather than prescribed as a constant \cite{schiehlen2004three}. Nevertheless, motions of the struck structure can cause the impactor to lose and regain contact several times during what would otherwise appear to be a single impact \cite{stoianovici1996critical}. Whether the impactor separates permanently after the initial contact, remains in sustained contact, or undergoes repeated contacts before eventually separating or sticking is thus determined by the dynamics of the coupled system.

For an infinite plate with Hertzian contact, the impact response can be described by a single nondimensional parameter \cite{zener1941intrinsic,olsson1992impact}. A related problem was studied in \cite{BHATTACHARJEE2018147}, who considered the transverse impact of a compact rigid body on an initially stationary, undamped, infinitely long Euler-Bernoulli beam, with contact governed by a Hertzian law. The beam response led to a fractional-order differential equation for the contact-point displacement, coupled to the nonlinear dynamics of the impacting body through the Hertzian contact force. The resulting system was solved numerically using a numerical recipe for fractional-order equations \cite{das2013simple}. The contact dynamics were governed by a single nondimensional parameter, denoted by $S$ in \cite{BHATTACHARJEE2018147}. For small $S$, the impactor separated permanently after the initial contact, whereas for large $S$, sustained contact occurred without separation. At intermediate values, repeated contact and separation events were possible before sustained contact eventually developed, with the number of separations increasing sharply over a narrow range of $S$. The study also provided a semi-analytical approximation for the initial contact phase and an asymptotic description of the long-time sustained-contact response. Interestingly, the latter was found to be independent of the preceding contact and separation history.

The Euler-Bernoulli beam model, however, neglects cross-section wise rotary inertia and consequently predicts an unbounded wave speed at high frequencies. This limitation is particularly relevant to impact, where the contact duration is short and the excitation contains substantial high-frequency content. It is reflected directly in the impulse response at the struck point, which for an Euler-Bernoulli beam grows as $\sqrt{\tau}$ at short times, where $\tau$ denotes time. Thus, an impulsive force produces an unbounded initial velocity at the contact point. The Rayleigh beam model retains the kinematics of the Euler-Bernoulli theory while accounting for rotary inertia, thereby providing bounded wave speeds \cite{rayleigh1894theory,timoshenko1930strength,meirovitch1967analytical,graff2012wave,whitham2011linear}. Its impulse response instead grows linearly with time at short times and approaches the $\sqrt{\tau}$ dependence at long times \cite{chatterjee2004short,biswas2026transverse}. The contact point therefore initially responds with a finite velocity, with rotary inertia dominating the short-time dynamics, before the long-time behaviour characteristic of the Euler-Bernoulli beam emerges. Rotary inertia thus introduces an intrinsic time scale into the problem. The Rayleigh beam is consequently more appropriate for studying brief impact interactions while retaining a relatively simple mathematical description. Although shear deformation is still neglected (more on which below), this model removes the unbounded wave speed and admits a closed-form impulse response \cite{biswas2026transverse}.

Impact problems involving beams with bounded wave speeds have been studied previously, primarily using the Timoshenko beam theory, which accounts for both rotary inertia and shear deformation\footnote{%
	Interestingly, allowing shear deformation makes the impulse response of the Timoshenko beam at the struck point show an instantaneously nonzero displacement, which is absent from the impulse response of the Rayleigh beam \cite{biswas2026transverse}.}.
These studies have considered transient impact responses of finite and infinite beams, including Hertzian contact, prescribed impact loading, and wave-front descriptions of the interaction \cite{park1994transient,zhang2021elastic,rossikhin1995ray,rossikhin1996impact,rossikhin2016impact,adamek2018comparison,szalai2013impact}. Comparisons between Euler-Bernoulli and Timoshenko beam models have also addressed the regularity of the point-contact problem \cite{szalai2013impact}. However, the contact dynamics of a Hertzian body impacting an infinite beam with bounded wave speed have not, to our knowledge, been systematically classified. In particular, the occurrence of permanent separation, sustained contact, and repeated contact between these two limiting behaviours, as studied for the Euler-Bernoulli beam in \cite{BHATTACHARJEE2018147}, remains unexplored. The present work follows up on \cite{BHATTACHARJEE2018147} by investigating the significant ways in which the interaction changes when rotary inertia is included through the Rayleigh beam model.

We emphasize that the Rayleigh beam impact problem is not a straightforward extension of the Euler-Bernoulli problem. The essential difference lies in the time dependence of the beam impulse response. Whereas the Euler-Bernoulli impulse response is homogeneous in time, allowing the coupled impact equations to be reduced to a fractional-order system \cite{BHATTACHARJEE2018147,das2013simple}, the Rayleigh beam impulse response contains a different initial regime due to rotary inertia, followed by a long term behaviour that approaches the Euler Bernoulli behaviour; in particular, it is not homogeneous in time. The resulting impact problem therefore cannot be reduced to the same fractional-order formulation, nor can this inhomogeneous time response be easily handled by rescaling time and displacement. Consequently, the single nondimensional parameter $S$ governing the Euler-Bernoulli problem is replaced by two independent parameters, denoted by $\Gamma$ and $\Lambda$, which are defined in due course below. The beam displacement at the contact point is obtained by convolving the dynamic contact force with the Rayleigh beam impulse response, which is available in both spectral integral and closed forms \cite{biswas2026transverse}. This gives a natural formulation of the coupled impact problem, but also introduces a computational difficulty: direct evaluation of the convolution at every time step requires $O(\tau^2)$ operations for a simulation extending up to time $\tau$ \cite{singh2006galerkin,singh2010beyond}. Such computation is prohibitive for the long-time simulations needed to distinguish permanent separation from eventual sustained contact, and for a systematic study over the two-dimensional $(\Gamma,\Lambda)$ parameter space.

One might be tempted to consider the problem as being governed by a nonlinear version of the Volterra integral equation of the second kind, given that numerical methods for Volterra equations and irrational fractional-order differential equations are well established \cite{brunner2004numerical,tang2008spectral,katani2015block,das2013simple}. However, nonlinearity and intermittent contact wherein the separation times are not known {\em a priori}
create significant difficulties. 

The approach we prefer is to approximate the infinite beam's contact point response using a finite-dimensional, time-invariant system, such that simulation upto time $\tau$ requires ${\cal O}(\tau)$ computations. Note that, in this approach, we approximate the dynamical system rather than any individual solution thereof. The principle is well known, as noted below. A significant contribution here lies in our variable transformation within the spectral integral which makes the required order of the approximating system smaller for the same level of accuracy.

Broadly speaking, finite-dimensional approximations of convolution operators can be constructed using diffusive representations of fractional operators \cite{montseny1998diffusive,montseny2004simple}, quadrature of the diffusive integral \cite{yuan2002numerical}, and Galerkin projection \cite{singh2006galerkin}. The Galerkin approach was subsequently extended from fractional operators to a broader class of convolution kernels \cite{singh2010beyond}. Here we adapt that framework to the Rayleigh beam impulse response, with a transformation of the frequency variable within the spectral integral into an abstract variable  such that, with only $N=25$ retained states with first order dynamics, the impulse response is reproduced to within $0.1\%$ relative error over fourteen decades in time. The resulting impact solutions agree closely with those obtained by direct convolution, but the computation time (for our parameter study below) is lower by about two orders of magnitude.

The contributions of this paper are, therefore, threefold. First, following up on the Euler-Bernoulli impact problem studied in \cite{BHATTACHARJEE2018147}, we investigate Hertzian impact on an infinite Rayleigh beam, for which rotary inertia introduces a second independent nondimensional parameter. Second, we develop a reduced-order model based on Galerkin projection in an abstract domain, which replaces the convolution-based beam dynamics with a moderately-sized system of local-in-time ODEs. Third, using this model, we classify the $(\Gamma,\Lambda)$ parameter plane into three distinct impact regimes: (i) at one end, a single contact followed by permanent separation; (ii) at the other end, a single sustained contact; and (iii) an intermediate regime involving repeated contact and separation events of diminishing amplitudes, eventually followed by sustained contact. We further derive short- and long-time asymptotic descriptions of the boundary between permanent separation and eventual sustained contact. At long times, $\Gamma$ and $\Lambda$ combine into a single effective parameter equivalent to $S$, recovering the asymptotic structure of \cite{BHATTACHARJEE2018147}, with the corresponding regime boundary scaling as $\Gamma_c \propto \Lambda^{4/5}$. At short times, the beam behaves effectively as an immovable wall, and the boundary instead scales as $\Gamma_c \propto \Lambda$. Numerical simulations show a transition between these two asymptotic scalings, with the boundary evolving from the large-time scaling at small $\Lambda$ towards the short-time scaling at large $\Lambda$.

Our analysis below assumes a Hertzian sphere striking an initially stationary, undamped, infinitely long Rayleigh beam. Note that our analysis also applies to finite beams provided boundary reflections do not return to the impact point during the contact process \cite{biswas2026transverse}, i.e., provided the impact interaction is sufficiently brief.

The remainder of the paper is organised as follows. Section \ref{contspec}  quickly recalls the impulse response of the infinite Rayleigh beam from \cite{biswas2026transverse} and its short- and long-time asymptotic behaviour. Section \ref{impact} formulates the impact problem and introduces the nondimensional parameters $\Gamma$ and $\Lambda$. Section \ref{galfinsec} develops the modified Galerkin formulation and presents its numerical validation. Section \ref{paragl} classifies the $(\Gamma,\Lambda)$ plane into the three impact regimes. Section \ref{secboundary} derives the short- and long-time asymptotic descriptions and the corresponding regime-boundary scalings, and compares these predictions with the numerically computed boundary. Finally, Section \ref{conc} concludes the paper.

\section{Infinite Rayleigh beam impulse response}
\label{contspec}

For ease of reading, we first briefly recall relevant results for the impulse response of an infinite Rayleigh beam from \cite{biswas2026transverse}. Consider an infinite Rayleigh beam with flexural rigidity $EI$, mass per unit length $\rho A$, and rotary inertia per unit length $\rho I$. Then the transverse displacement $u(x,t)$ at any point $x$ satisfies
\begin{equation}
	EI\frac{\partial^4 u}{\partial x^4}
	+\rho A\frac{\partial^2 u}{\partial t^2}
	-\rho I\frac{\partial^4 u}{\partial x^2\partial t^2}
	=
	P\delta(x)\,\delta(t), \qquad
	u\,(x,0^-)=0, \quad v\,(x,0^-)=\partial u / \partial t |_{t=0^-}=0,
	\label{eq1}
\end{equation}
where a transverse impulse of magnitude $P$ is applied at $x=0$ and $t=0$ on a beam that starts at rest. We nondimensionalize using
\begin{equation}
	\bar u=u\sqrt{\frac{A}{I}},
	\quad
	\bar P=\frac{P}{\sqrt{EA\rho I}},
	\quad
	\tau=t\sqrt{\frac{EA}{\rho I}},
	\quad
	\bar x=x\sqrt{\frac{A}{I}},
	\quad
	\bar b=b\sqrt{\frac{A}{I}},
	\quad
	\bar L=L\sqrt{\frac{A}{I}},
	\quad
	\bar v=v\sqrt{\dfrac{\rho}{E}},
	\label{eq1nondscale}
\end{equation}
obtaining
\begin{equation}
	\frac{\partial^4\bar u}{\partial\bar x^4}
	+\frac{\partial^2\bar u}{\partial\tau^2}
	-\frac{\partial^4\bar u}{\partial\bar x^2\partial\tau^2}
	=
	\bar P\delta(\bar x)\,\delta(\tau).
	\label{eq1nond}
\end{equation}
The impulse response function at the struck point is (setting $\bar P = 1$)
\begin{equation}
	\bar G(\tau)
	=\frac{1}{\pi} \displaystyle
	\int_0^\infty
	\frac{
		\sin\left(
		\dfrac{ \kappa ^2}{\sqrt{1+ \kappa ^2}}\tau
		\right)
	}{
		\kappa ^2\sqrt{1+ \kappa ^2}
	}
	\, \mathrm d \kappa  
	=
	\frac{\tau}{2}e^{-\tau}
	\left[I_0(\tau)+I_1(\tau)\right],
	\label{eqG}
\end{equation}
where $I_n(\cdot)$ denotes the modified Bessel function of the first kind.
The short-time expansion of $\bar G(\tau)$ is
\begin{equation}
	\bar G(\tau)
	=
	\frac{\tau}{2}
	-\frac{\tau^2}{4}
	+\frac{\tau^3}{8}
	-\frac{5\tau^4}{96}
	+\mathcal{O}(\tau^5),
	\qquad
	\tau\rightarrow0,
	\label{eqshortcont}
\end{equation}
whereas the large-time expansion is
\begin{equation}
	\bar G(\tau)
	=
	\frac{\sqrt{\tau}}{\sqrt{2\pi}}
	\left[
	1-\frac{1}{8\tau}
	-\frac{3}{128\tau^2} -\dfrac{15}{1024 \tau^3}
	+\mathcal{O}(\tau^{-4})
	\right],
	\qquad
	\tau\rightarrow\infty.
	\label{eqlongcont}
\end{equation}

\section{Hertzian impact formulation}
\label{impact}


\subsection{Dimensional equations}
\label{secdim}

\begin{figure}[htbp]
	\centering
	\includegraphics[width=0.45\textwidth]{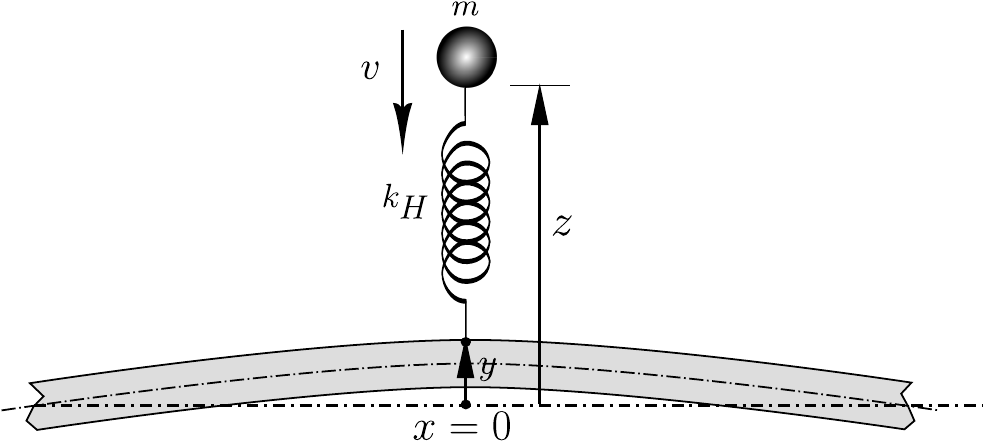}
	\caption{A spherical Hertzian impactor of mass $m$ strikes an infinite Rayleigh beam at $x=0$ with initially downward velocity $v$, with contact mediated by a Hertzian spring of stiffness $k_H$. The size of the spring is exaggerated: in reality it is microscopic. The beam has flexural rigidity $EI$, mass per unit length $\rho A$, and rotary inertia per unit length $\rho I$. Displacements are positive upwards. Rebound of the impactor implies that its post-impact velocity is upwards.}
	\label{fig:hertzscheme}
\end{figure}

See Fig.\ \ref{fig:hertzscheme}. Consider a spherical impactor of mass $m$ striking an infinite Rayleigh beam at $x=0$. Gravity is neglected. We consider dimensional quantities first, and will nondimensionalize later.

Let $y(t)$ and $z(t)$ denote the transverse displacements of the notional contact points on the beam and the sphere, respectively. The indentation is
\begin{equation}
	\label{gg11a}
	\eta(t)=y(t)-z(t),
\end{equation}
and the Hertz contact force $f(t)$ is 
\begin{equation}
	f(t)=k_H\eta(t)^{3/2}H(\eta(t)), \label{impact_force_dim}
\end{equation}
where $k_H$ is the Hertz contact stiffness and $H(\cdot)$ is the Heaviside function,
\[H(\eta) = 1 \mbox{ if } \eta \ge 0, \mbox{ and  zero otherwise}.\]
The beam is initially stationary and the impactor approaches with velocity $v$.

The nondimensionalized
impulse response function
$\bar G(\tau)$ was presented in Eq.\ \eqref{eqG}. Let the dimensional equivalent, also for a unit impulse, be $G(t)$. As the dynamics evolves, with time-varying contact force
$f(t)$, the response
\[y(t) = - \int_0^t f(\xi) G(t -\xi) \, {\rm d}\xi,\]
which is the source of the convolution integral in the dynamics (the minus sign is because the contact force acts downward while $y$ is measured upward). The full governing equations are   
\begin{equation}
	\begin{aligned}
		&m\ddot z(t)=f(t),
		\qquad
		y(t)=-\int_0^t f(\xi)G(t-\xi)\,\mathrm d\xi, \\
		&\text{with} \qquad
		z(0)=0,
		\qquad
		\dot z(0)=-v, \qquad \mbox{and } y=0 \mbox{ automatically},	\end{aligned}
	\label{impact_dim}
\end{equation}
where overdots denote derivatives with respect to the dimensional time $t$, and $G(t)$ is the dimensional impulse response function at the point of contact. The system is nonlinear because of Eqs.\ \eqref{gg11a} and
\eqref{impact_force_dim}.
We know physically that $\dot y(0) = 0$, and that also emerges through $G(t)$. 


\subsection{Nondimensional equations}
\label{secnondim}
We nondimensionalize Eqs.\ \eqref{impact_dim} using Eq.\ \eqref{eq1nondscale}. Additionally, we now have the nondimensional impacting mass $$\bar m = m / (\rho \sqrt{AI}).$$ Equations \eqref{impact_dim} yield
\begin{equation}
	\begin{aligned}
		&\bar z''=\lambda_0\bar F,
		\qquad
		\bar y=-\gamma_0
		\int_0^\tau
		\bar F(\zeta)\,\bar G(\tau-\zeta)\,\mathrm d\zeta,\\
		&\text{where}\qquad
		\bar F=\bar\eta^{3/2}H(\bar\eta),
		\qquad
		\bar\eta=\bar y-\bar z,\qquad \bar y(0)=\bar y'(0)=\bar z(0)=0,
		\qquad
		\bar z'(0)=-\bar v,\\
		&\qquad \lambda_0=
		\dfrac{k_H}{\bar m EA}
		\left(\dfrac{I}{A}\right)^{3/4},
		\qquad \text{ and } \qquad
		\gamma_0=
		\dfrac{k_H}{EA}
		\left(\dfrac{I}{A}\right)^{3/4}
		=\bar m\lambda_0,
	\end{aligned}
	\label{eq:nondim_impact}
\end{equation}
and where primes denote derivatives with respect to the nondimensional time $\tau$ and $\bar G$ is the nondimensional impulse response function. This nondimensionalization gives three dimensionless terms: $\bar v$, $\lambda_0$ and $\gamma_0$, which satisfy $\gamma_0=\bar m\lambda_0$. These three terms can further be reduced by normalizing the already nondimensionalized displacements with $\bar v$ as
\begin{equation}
	\bar y=\bar v\,Y(\tau),
	\qquad
	\bar z=\bar v\,Z(\tau).
	\label{eq:amprescale}
\end{equation}
Equations \eqref{eq:nondim_impact} then become (introducing some new symbols to avoid abuse of notation)
\begin{equation}
	\begin{aligned}
		&Z''=\Lambda F,
		\qquad
		Y=-\Gamma\int_0^\tau
		F(\zeta)\,\bar  G(\tau-\zeta)\,\mathrm d\zeta,\\
		&\text{where}\qquad
		F=\alpha^{3/2}H(\alpha),
		\qquad
		\alpha=Y-Z,
		\qquad
		Y(0)=Y'(0)=Z(0)=0,
		\qquad
		Z'(0)=-1,\\
		&\text{and}\qquad
		\Lambda=\sqrt{\bar v}\,
		\dfrac{k_H}{\bar m EA}
		\left(\dfrac{I}{A}\right)^{3/4},
		\qquad
		\Gamma=\sqrt{\bar v}\,
		\dfrac{k_H}{EA}
		\left(\dfrac{I}{A}\right)^{3/4}.
	\end{aligned}
	\label{eq:nondfinal_impact}
\end{equation}
We see that, after the velocity normalization, the impact problem is governed by the two dimensionless parameters $\Lambda$ and $\Gamma$. The dimensionless mass $\bar m$ is retained through the relation $\Lambda=\Gamma/\bar m$. We note that in the Euler-Bernoulli case, an additional time scaling reduces the problem to one with a single dimensionless parameter \cite{BHATTACHARJEE2018147}; but here we have two.

\section{Finite dimensional approximation of the impact dynamics}
\label{galfinsec}

The beam displacement in Eq.\ \eqref{eq:nondfinal_impact} is expressed as a convolution of the contact force $F$ with the impulse response function $\bar G$, making the formulation nonlocal in time. We replace this representation with a finite dimensional Galerkin approximation of the beam dynamics, yielding a system of local-in-time ordinary
differential equations coupled to the impactor equation through the Hertz contact force. Note that the Galerkin approximation does {\em not} use a modal expansion for the beam, because an infinitely long beam does not have normal modes.

\subsection{Galerkin formulation {\em via} an abstract variable}
\label{galerkin}

We adapt the Galerkin projection framework described in \cite{singh2006galerkin,singh2010beyond}, but after transforming to a new abstract variable which gives more rapid convergence in the reduced order model. This transformation is new to the best of our knowledge, especially for such an application; and it makes the computation simpler and quicker. 
We begin with the Laplace transform of
$\bar G$, and transform the Laplace variable to another abstract variable.

The Laplace transform from Eq. \eqref{eqG}  (see \cite{biswas2026transverse}) is
\begin{equation}
	\mathcal{L}\{\bar G\}
	=
	\frac{1}{2s\sqrt{s(s+2)}}.
	\label{eq:lap}
\end{equation}
Multiplying by $s$ gives $\displaystyle s\mathcal{L}\{\bar G\}
=
\frac{1}{2\sqrt{s(s+2)}}$. We use the identity (choosing the principal square root branch $\sqrt{1-b^2/a^2}$)
\[ \int_0^\pi\frac{\mathrm{d}\theta}{a-b\cos\theta}
=
\frac{\pi}{a\sqrt{1-b^2/a^2}},\quad a,b \in \mathbb{C},\]
where $a-b\cos\theta\neq0$ for all $\theta\in[0,\pi]$, or equivalently, $a/b\notin[-1,1]$ for $b\neq0$; and $a\neq0$ for $b=0$. With $a=s+1$ and $b=1$ (which satisfy the conditions on $a$ and $b$), Eq.~\eqref{eq:lap}
can be written as
\begin{equation}
	s\mathcal{L}\{\bar G\}
	=
	\frac{1}{2\pi}
	\int_0^\pi
	\frac{\mathrm{d}\theta}{s+1-\cos\theta}.
\end{equation}
In this way, we have first replaced the explicitly known function of Eq.~\eqref{eq:lap} with a definite integral.
Now introducing $\mu=1-\cos\theta,
$ for which $\displaystyle \mathrm{d}\theta=\frac{\mathrm{d}\mu}{\sqrt{\mu(2-\mu)}}$, we have
\begin{equation}
	\mathcal{L}\{\bar G\}
	=
	\int_0^2
	\frac{\mathrm{d}\mu}{2\pi s\, (s+\mu)\sqrt{\mu(2-\mu)}}.
\end{equation}
Towards inverting the Laplace transform, but exchanging the order of inversion and integration, and using
\[\displaystyle \mathcal{L}^{-1}\left\{\frac{1}{s(s+\mu)}\right\}
=
\frac{1-e^{-\mu\tau}}{\mu},\]
we obtain a new integral form of the impulse response function,
\begin{equation}
	\bar G(\tau)
	=
	\int_0^2
	\frac{1-e^{-\mu\tau}}
	{2\pi\mu\sqrt{\mu\,(2-\mu)}}\,\mathrm{d}\mu.
	\label{eqGlam}
\end{equation}
Note that Eq.~\eqref{eqGlam} is exactly equivalent to Eq.\ \eqref{eqG}. The intervening Laplace transform calculations can now be set aside.
Working with Eq.~\eqref{eqGlam}, we now reintroduce $\mu=1-\cos\theta$ and obtain
another equivalent expression,
\begin{equation}
	\bar G(\tau)
	=
	\frac{1}{2\pi}\int_0^\pi
	\frac{1-e^{-(1-\cos\theta)\tau}}
	{1-\cos\theta}\, \mathrm{d}\theta.
	\label{eqGtheta}
\end{equation}
In the above, the apparent singularity at $\theta=0$ is removable. Since $1-e^{-(1-\cos\theta)\tau}
\sim
(1-\cos\theta)\tau$, the integrand tends to $\tau$ as $\theta\to0$ and remains bounded throughout the interval. Now, we define the abstract variable $q(\theta,\tau)=e^{-(1-\cos\theta)\tau}$, $\tau\geq0$, which satisfies
\begin{equation}
	\frac{\partial q}{\partial\tau}+(1-\cos\theta)\,q=\delta(\tau),
	\qquad
	q(\theta,0^-)=0,
	\label{eqdecay}
\end{equation}
where $\delta(\tau)$ denotes the Dirac delta function. So, $q$ is the response of a first-order system in $\tau$, with $\theta$ as a parameter, to a unit impulse. Eq.~\eqref{eqGtheta} can then be written as
\begin{equation}
	\bar G(\tau)
	=
	\frac{1}{2\pi}\int_0^\pi
	\frac{1-q(\theta,\tau)}{1-\cos\theta}\,\mathrm{d}\theta,
	\qquad \tau\geq0.
	\label{eqGthetadec}
\end{equation}
Equivalently, the time derivative of $\bar G$ can be constructed from $q$ by
\begin{equation}
	\bar G'(\tau) =\frac{1}{2\pi}\int_0^\pi q(\theta,\tau)\,\mathrm{d}\theta, \qquad \bar G(0)=0.
	\label{eqGthetadecn}
\end{equation}
In this sense, using either Eq.\
\eqref{eqGthetadec} or \eqref{eqGthetadecn}, the integrated output of the abstract $q$-system provides the same impulse response function $\bar G$, driven by the same unit impulse.

We are finally ready to develop a finite dimensional approximation. We approximate the $\theta$-dependence of $q$ in Eq.~\eqref{eqdecay}
using a finite basis and Galerkin projection, and reconstruct the approximate
impulse response function $\bar G_N$ from the Galerkin approximation of $q$.

The long-time $\sqrt{\tau}$
growth of $\bar G$ is in the neighbourhood of $\mu=0$, {\it i.e.},
$\theta=0$. To resolve this region more accurately with fewer retained basis
functions ({\it i.e.}, in the interest of a lower order model), we introduce
the stretched coordinate
\begin{equation}
	\theta=\pi \beta^{2},\qquad \beta\in[0,1],
	\label{eq:betadef}
\end{equation}
and use the cosine basis
\begin{equation}
	\phi_0(\beta)=1,\qquad
	\phi_n(\beta)=\sqrt2\cos(n\pi \beta),\quad n=1,2,\ldots, N-1.
\end{equation}
However, our Galerkin projection retains the original integral measure
$\mathrm{d}\theta$, which becomes $J(\beta)\,\mathrm{d}\beta$ under the
mapping of Eq.\ \eqref{eq:betadef}, with $$J(\beta)=2\pi\beta.$$
We therefore
define the inner product
\begin{equation}
	\langle f,g\rangle
	=\frac{1}{\pi}\int_0^1 f(\beta)g(\beta)J(\beta)\,\mathrm{d}\beta.
	\label{eqinpro}
\end{equation}
Approximating $q$ as
\begin{equation}
	q_N(\beta,\tau)=\sum_{j=0}^{N-1}a_j(\tau)\phi_j(\beta),
	\label{eqtri}
\end{equation}
and imposing Galerkin orthogonality on Eq.~\eqref{eqdecay}, with $q$ replaced by $q_N$,
gives
\begin{equation}
	\mathbf M{\mathbf a}'+\mathbf K\mathbf a=\mathbf g\,\delta(\tau),
	\qquad
	\mathbf a(0^-)=\mathbf 0,
	\label{eqgal}
\end{equation}
where $\mathbf a=(a_0,\ldots,a_{N-1})^T$ are the Galerkin coefficients (and generalized coordinates), and
\begin{equation}
	M_{ij}=\langle\phi_i,\phi_j\rangle,
	\qquad
	K_{ij}=\langle\phi_i,\mu\phi_j\rangle,
	\qquad
	g_i=\langle\phi_i,1\rangle,
	\quad
	\mbox{ where } 
	\mu(\beta)=1-\cos(\pi\beta^2).
\end{equation}
Equation \eqref{eqgal} is a standard linear system of ODEs. The impulse gives $\mathbf a(0^+)=\mathbf M^{-1}\mathbf g$. The matrices are evaluated using Gauss-Legendre quadrature on $[0,1]$.
Replacing $q$ by $q_N$ in Eq.~\eqref{eqGthetadecn} gives
the approximation
\begin{equation}
	\bar G_N'
	=\frac{1}{2\pi}\int_0^1 q_N(\beta,\tau)J(\beta)\,\mathrm{d}\beta
	=\mathbf c^T\mathbf a(\tau),
	\qquad
	\mathbf c=\tfrac12\mathbf g,
	\qquad
	\bar G_N(0)=0.
	\label{eq:dgndtau}
\end{equation}

Our finite-dimensional memoryless (local-in-time) model for the impact problem in Eq.~\eqref{eq:nondfinal_impact} is now ready. Since the beam is loaded by the
contact force $F(\tau)$ instead of a unit impulse, and $Y$ is the convolution of
$\bar G$ with $F$, our impact model becomes
\begin{equation}
	\begin{aligned}
		&F=(Y-Z)^{3/2}H(Y-Z),\\
		&Z''=\Lambda F, \qquad
		\mathbf M{\mathbf b}'+\mathbf K\mathbf b=\mathbf g\,F, \qquad Y'=-\Gamma\,\mathbf c^T\mathbf b, \\
		&Y(0)=0,\quad Z(0)=0,\quad Z'(0)=-1,\quad \mathbf b(0)=\mathbf 0,
	\end{aligned}
	\label{eq:nondfinal_impact_galerkin}
\end{equation}
where we recall that $H(.)$, the Heaviside function, accounts for making and breaking of contact during the impact interaction.
The system is nonlinear because $F$ is nonlinear. Note that this approximated system has $N+3$ states, namely $Z$, $Z'$, $Y$, and $b_0,\ldots,b_{N-1}$;
and it can be integrated directly with any standard ODE solver. This compact, finite dimensional, approximate numerical model for the impact interaction derives its high accuracy from the $s$-$\theta$-$\beta$ transformations used above, and is an important contribution of this paper.

\subsection{Numerical verification}
\label{numval}

\begin{figure}[!htbp]
	\centering
	\includegraphics[width=\textwidth]{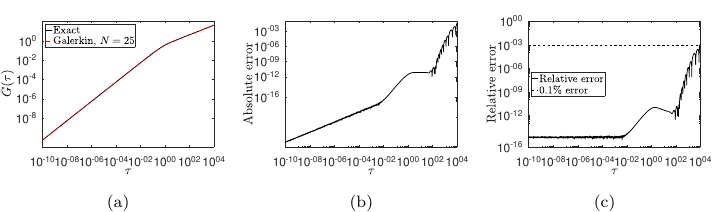}
	\caption{Approximated and exact impulse responses:
		(a) impulse response, (b) absolute error, and (c) relative error. The noise floor in (c) corresponds to our use of double precision (16 digits).}
	\label{fig:gcomb}
\end{figure}

The accuracy of the Galerkin approximation is first assessed by comparing the
approximated impulse response function, namely $\bar G_N$, with the exact impulse response function
of Eq.~\eqref{eqG}. Note that, in approximating the impulse response function, we are approximating the dynamic system and not a specific solution.
Figure~\ref{fig:gcomb} shows the comparison over $10^{-10}\leq\tau\leq10^4$ for
$N=25$. The relative error remains below $0.1\%$ throughout this range and is
substantially smaller over most of it. Larger $N$ would give us even greater accuracy. We use $N=25$ in all subsequent
impact calculations.

The time integration of the Galerkin projection based system for approximating the impulse response was performed using the
\texttt{lsode} solver in GNU Octave, with relative and absolute tolerances of
$10^{-10}$ and $10^{-12}$, respectively. The relevant system matrices  were evaluated using $400$-point
Gauss-Legendre quadrature (in numerical experiments, no meaningful changes occurred upon increasing the number  beyond 400 for $N=25$). 

The nonlinear impact problem solution was also checked by comparing the Galerkin solution of
Eq.~\eqref{eq:nondfinal_impact_galerkin} with a direct numerical solution of the
convolution-based formulation of Eq.\ \eqref{eq:nondfinal_impact}, for $\Gamma = 700$ and $\Lambda = 0.1$. The direct
convolution-based solution is not offered here as an independent benchmark because its errors have not been quantified, and it is computationally very expensive as well. We do {\em not} claim that it is more accurate than the Galerkin-based solution. However, it is accurate to more than plotting accuracy, and is included to check that the two independent and quite different approaches
agree. This check provides an assurance that there is no inadvertent algorithmic error in the Galerkin-based solution.

\begin{figure}[htbp]
	\centering
	\includegraphics[width=\textwidth]{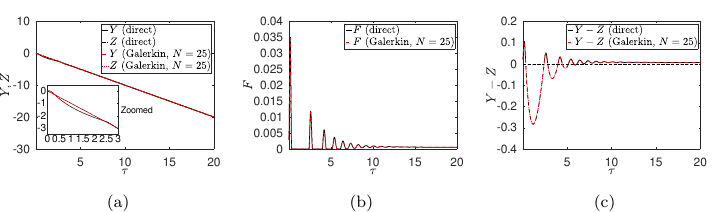}
	\caption{Comparison of the direct convolution and Galerkin solutions for the
		Hertzian impact problem: (a) beam and impactor displacements, (b) contact
		force, and (c) contact indentation. Here, $\Gamma = 700$ and $\Lambda = 0.1$. The two approaches agree to better than plotting accuracy. Observe that the impact ends with sustained contact (more on this below).}
	\label{figimpcomp}
\end{figure}

In the direct calculation, using the exact impulse response $\bar G(\tau)$ of
Eq.~\eqref{eqG}, the convolution integral was evaluated by composite Simpson
quadrature\footnote{%
	The $1/3$ rule for an even number of intervals, and the $3/8$ rule in cases where there are 3 intervals left at the end.}.
At each time step, the contact force was found by a relaxed
fixed-point iteration, started from a linear extrapolation of the previous two
force values. The impactor displacement was advanced using the trial contact
force, and its velocity was updated using the trapezoidal rule for the
acceleration. We used a time step $\Delta\tau = 0.02$ over $0 \leq \tau \leq 20$.
The iteration at each step was terminated when the relative change in the contact
force fell below $10^{-4}$, or after a maximum of $20$ iterations. As emphasized above, the aim of this convolution-based solution was comparison only, and the above algorithm was sufficient for plotting accuracy. The Galerkin based solution was much more accurate.

Figure~\ref{figimpcomp} shows satisfactory agreement between the two different solutions for the
beam and impactor displacements, the contact force, and the indentation. This
example involves repeated contact and separation, and the Galerkin solution
captures it well. We use only the Galerkin formulation for the parameter study in the next section.


\section{Impact regime classification}
\label{paragl}

We now numerically examine the impact dynamics over the $\Gamma$-$\Lambda$ parameter space. Figures~\ref{figcontactstartend} and \ref{figcontactstartenda} show the contact start and end times as functions of $\Gamma$ for twelve values of $\Lambda$, with $10^{-1}\leq\Gamma\leq10^3$. Each sweep uses 1251 equally spaced values of $\Gamma$. The integration is restricted to $0\leq\tau\leq1000$ due to computational limitations (our use of $N=25$; recall Figure \ref{fig:gcomb}(c)).

\begin{figure}[!htbp]
	\centering
	\includegraphics[width=\textwidth]{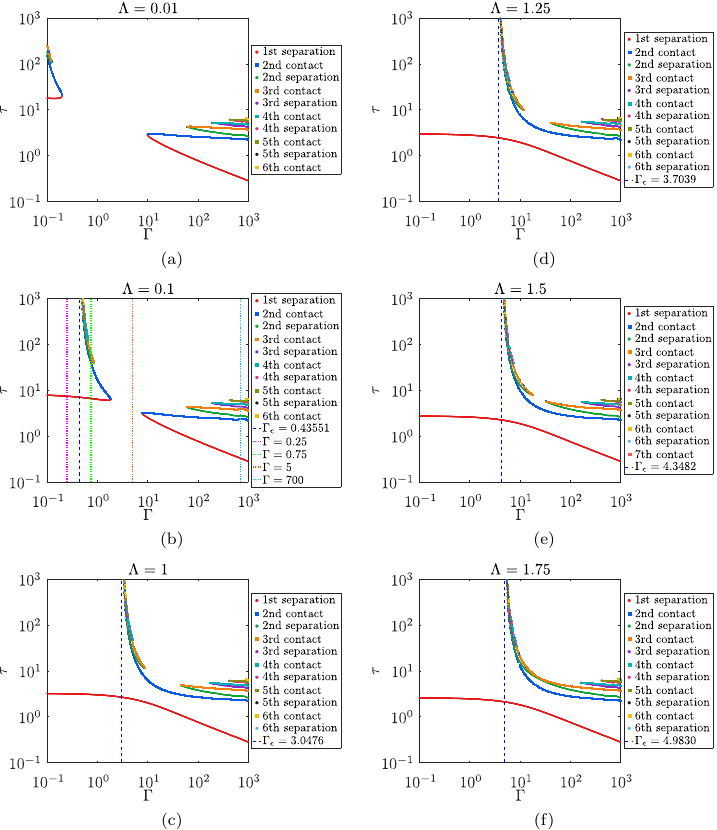}
	\caption{Contact start and end (separation) times as functions
		of $\Gamma$ for (a) $\Lambda=0.01$, (b) $\Lambda=0.1$,
		(c) $\Lambda=1$, (d) $\Lambda=1.25$, (e) $\Lambda=1.5$,
		and (f) $\Lambda=1.75$, over $10^{-1}\leq\Gamma\leq10^3$.
		Both axes are logarithmic. The vertical dashed line in each panel marks the smallest
		$\Gamma$ for which a second contact is observed within the
		simulated time interval. For $\Lambda=0.01$ and $0.1$ (a, b),
		the gaps indicate parameter ranges in which no separation occurs
		after first contact. In panel (b), the additional vertical lines
		indicate five values of $\Gamma$ ($0.25$, $0.4831$, $0.75$,
		$5$, and $700$) which will be considered in another figure in due course.}
	\label{figcontactstartend}
\end{figure}

\begin{figure}[!htbp]
	\centering
	\includegraphics[width=\textwidth]{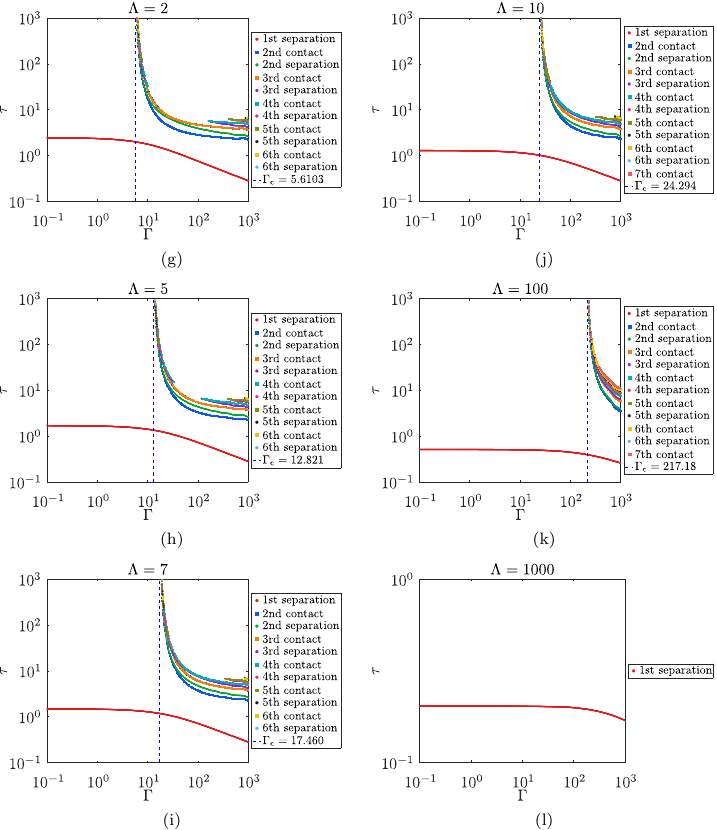}
	
	\caption{Continuation of results from Figure \ref{figcontactstartend}.
		Panels (g) $\Lambda=2$, (h) $\Lambda=5$, (i) $\Lambda=7$,
		(j) $\Lambda=10$, (k) $\Lambda=100$, and (l) $\Lambda=1000$.
		The first separation marks the end of the initial contact episode.
		Subsequent branches indicate the start and end of later contact
		episodes. }
	\label{figcontactstartenda} 
\end{figure}

The first contact begins at $\tau=0$. Recall
\[\alpha = Y-Z,\]
which denotes the indentation. Contact and separation times are identified from zero crossings of $\alpha$, located by linear interpolation between output samples spaced $\Delta\tau = 0.1$ apart for regular plotting and much smaller intervals for some precise estimates below. A separation is a transition of $\alpha$ from positive to nonpositive values, at which instant the contact force
$F$ vanishes. A subsequent contact is the reverse transition, from nonpositive to positive $\alpha$, at which $F$ becomes positive. To suppress spurious events caused by numerical noise near
$\alpha=0$, a zero crossing is registered as a separation only if
$|\alpha|>2\times10^{-5}$, which
corresponds to an $|F|$ threshold of about $8.9\times10^{-8}$ (nondimensional).

In order to understand Figs.~\ref{figcontactstartend} and \ref{figcontactstartenda} (to be viewed as a pair: panels (a)-(f) are in Fig.~\ref{figcontactstartend}, and panels (g)-(l) are in Fig.~\ref{figcontactstartenda}), it is useful to think of $\Gamma$ as being determined by the lateral dimensions of the beam. If the characteristic lateral dimension of the beam is $r$, say, then we can think in terms of $\Gamma \sim 1/\sqrt{r}$ (see Eq.\ \eqref{eq:nondfinal_impact}). This means that large $\Gamma$ corresponds to a thin beam, and small $\Gamma$ corresponds to a fat beam. Also, beam properties held constant, since $\Lambda = \Gamma/\bar m$, we can think of $\Lambda$ as being inversely proportional to the impactor mass.

Starting with Fig.~\ref{figcontactstartend}(a), wherein the small $\Lambda$ means a large impactor mass, we observe an intermediate regime of $\Gamma$ values where contact, once initiated, is sustained: separation does not occur, at least within the simulation window. For larger $\Gamma$ (thinner beam), contact is made and broken multiple times. This can be explained in that the impulse response function for the beam is proportional to $\tau$ for small $\tau$, which means that the initial behaviour of the contact point on the beam is like that of a small point mass. As a result of being struck by a large impactor mass, the small mass runs ahead while the impactor velocity is only slightly reduced; and eventually the beam slows down (because the impulse response function is proportional to $\sqrt{\tau}$ for large $\tau$) and another contact event occurs. The larger the $\Gamma$, the larger is the number of subsequent contact events. However, we also note multiple contact events for small $\Gamma$, which corresponds to a fatter beam! This aspect is better understood by looking at Fig.~\ref{figcontactstartend}(b), where the qualitative behaviour of the left side of the plot has moved into the plotting range.

In Fig.~\ref{figcontactstartend}(b), the regime of sustained contact has shrunk: this is not surprising because now $\Lambda$ is a bit larger, and so the impactor mass is less extreme. The right side of Fig.~\ref{figcontactstartend}(b) resembles that of Fig.~\ref{figcontactstartend}(a). The left side of Fig.~\ref{figcontactstartend}(b) suggests that there is a range of small $\Gamma$ values where separation occurs after one contact, and then further contact is not established. This is consistent with physical intuition: a ball impacting a fat-enough beam can indeed bounce off after a single contact with its velocity direction reversed. The various vertical lines in the figure correspond to $\Gamma$ values that will be reexamined a little later in this paper. The blue dashed vertical line shows the smallest $\Gamma$ within the studied range for which a second contact occurs: we call this $\Gamma_c$. Such a $\Gamma_c$ is seen in all remaining subplots of Fig.~\ref{figcontactstartend} as well as in all subplots of Fig.~\ref{figcontactstartenda}. These $\Gamma_c$ values were computed by iteratively adjusting $\Gamma$ until the impactor's velocity at the instant of first separation was zero\footnote{%
	If the velocity is upwards, then clearly no further contact occurs. And if it is downward, then the impactor moves down like time $\tau$ while the beam eventually moves like $\sqrt{\tau}$, so further contact is guaranteed.}.

The situation changes qualitatively as $\Lambda$ increases. In 
Fig.~\ref{figcontactstartend}(c)-(e), we find coalescence between left and right lobes in the graphs, such that there is no longer a regime with sustained contact without separation. In Fig.~\ref{figcontactstartend}(f), we find coalescence between another pair of lobes, providing two more curves that run continuously from left to right, indicating a regime of $\Gamma$ wherein sustained contact occurs after {\em two} separations.

Now looking at Fig.~\ref{figcontactstartenda}, we find two interesting trends. The first is a continuation of the previously seen trend, wherein more lobes from left and right coalesce as $\Lambda$ increases, showing larger numbers of intervening separations before eventual sustained contact. The second is an overall movement of these contact and separation curves to the right, {\it i.e.}, in the direction of increasing $\Gamma$. This trend may be understood in terms of holding $\Gamma$ fixed ({\it i.e.}, holding the beam properties fixed) and increasing $\Lambda$ ({\it i.e.}, making the impactor mass smaller). Clearly, for any beam properties and contact stiffness combination held fixed, a sufficiently small impactor mass will lead to rebound after a single contact. For a thinner beam (larger $\Gamma$), this mass must be smaller (larger $\Lambda$).

A simple qualitative conclusion that emerges from this numerical parameter study is that if a second contact phase is initiated, then sustained contact is inevitable!

It is also seen from these figures (Figs.~\ref{figcontactstartend} and \ref{figcontactstartenda}) that, $\Lambda$ held fixed, there is a critical $\Gamma_c$ below which there is permanent separation, and above which eventual sustained contact is inevitable. The vertical dashed blue lines in Figs~\ref{figcontactstartend} and \ref{figcontactstartenda} indicate
$\Gamma_c$ for each case (except in panels (a) and (l), where it lies outside the computation window). As seen in these figures, the contact-time graphs asymptotically approach this critical value from above, and the first recontact occurs after a long time for $\Gamma$ slightly exceeding $\Gamma_c$.
Increasing $\Lambda$ increases $\Gamma_c$.

The foregoing results suggest that, depending on the parameters $\Lambda$ and $\Gamma$, there are three impact regimes: (i) a single contact episode followed by permanent separation, (ii) sustained contact without separation, and (iii) one or more separations followed by sustained contact.

We now examine time histories within some individual impacts.

\begin{figure}[!htbp]
	\centering
	\includegraphics[width=\textwidth,height=0.95\textheight]{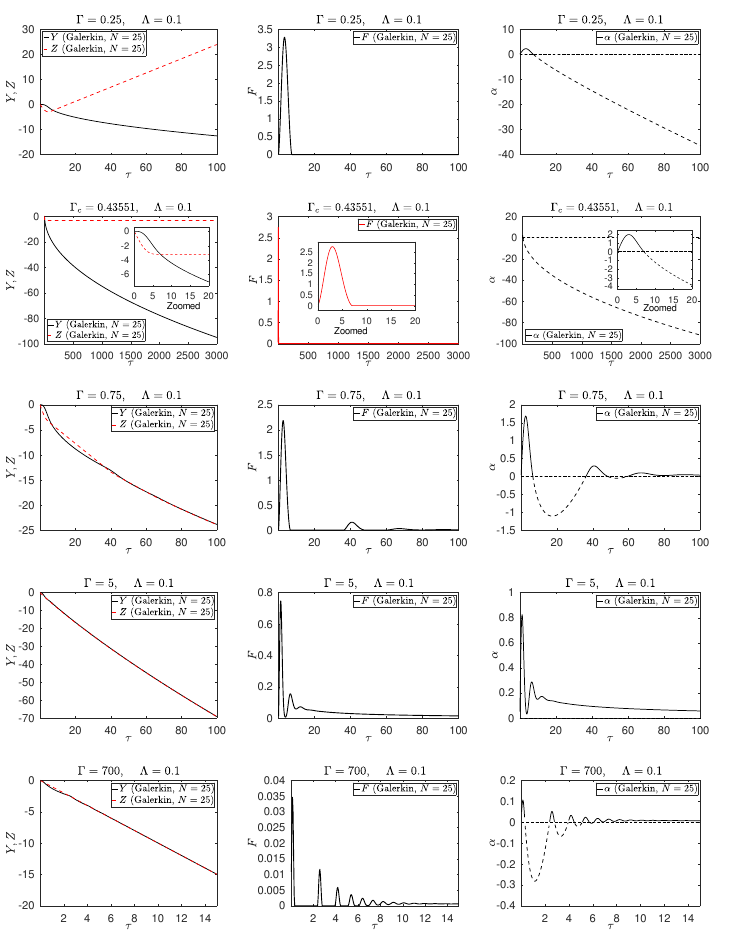}
	\caption{From left to right: $Y,Z$ vs $\tau$, $F$ vs $\tau$, and $\alpha$ vs $\tau$, respectively. From top to bottom: $\Gamma=0.25$, $0.43551$, $0.75$, $5$, and $700$, respectively. $\Lambda=0.1$ in all cases.}
	\label{fig:combined_gamma_cases}
\end{figure}

Figure~\ref{fig:combined_gamma_cases} shows the time histories for $\Lambda=0.1$ at the five increasing values of $\Gamma$ marked in Fig.~\ref{figcontactstartend}(b). For $\Gamma=0.25 < \Gamma_c$ (first row of Fig.~\ref{fig:combined_gamma_cases}), we have one contact followed by permanent separation. The displacement, contact force, and indentation plots all show this single contact event. 

For $\Gamma = 0.43551 = \Gamma_c$ (second row of Fig.~\ref{fig:combined_gamma_cases}), the impactor velocity at the end of first contact is zero as required (see the inset within the left subplot). Displacement, contact force, and indentation plots are all consistent with this physical intuition.

For $\Gamma=0.75 > \Gamma_c$ (third row of Fig.~\ref{fig:combined_gamma_cases}), we observe two separations preceding sustained contact as expected. In these three subplots, we see two phases where $Z<Y$ (left subplot), two relatively large contact force peaks (middle), and small but positive indentation for times after the second separation (dashed portions of the curve that dip below zero; right subplot).

Referring to Fig.~\ref{figcontactstartend}(b) again, for $\Gamma=5$, we expect sustained contact without separation. This is indeed observed in the fourth row of Fig.~\ref{fig:combined_gamma_cases}. The indentation remains positive, approaching zero from above, with the contact force undergoing several decaying oscillations as it approaches zero. 

Finally, for $\Gamma=700 \gg \Gamma_c$ (fifth row of Fig.~\ref{fig:combined_gamma_cases}), the response consists of many brief contact episodes with progressively smaller peaks, followed by sustained contact as the indentation goes to zero.

Across all these cases, successive local maxima in the contact force (wherever applicable) show decreasing values, consistent with an approach to sustained contact.

We now proceed to analytical treatment of some of the phenomena observed in this section.

\section{Asymptotic analysis}
\label{secboundary}
The impact system in Eq.~\eqref{eq:nondfinal_impact} contains two
independent dimensionless parameters, $\Gamma$ and $\Lambda$. Let $\tau_s$ denote a time such that, for $\tau>\tau_s$, there is either final separation or established sustained contact. In other words, all intervening separations and/or individual contact phases have occurred well before $\tau_s$. We refer to those intervening contact and separation events as the system's initial dynamics, and the behaviour after $\tau_s$ as the final dynamics.

The two separate regimes of $\tau_s\gg1$ and $\tau_s\ll1$ correspond to very
long and very short phases of initial dynamics, respectively. In these two
limiting regimes the kernel $\bar G(\tau)$ can be replaced by its corresponding
asymptotic form to leading order only, and in these regimes $\Gamma$ and $\Lambda$ combine into a single
parameter in each case. These regimes are treated separately below.


\subsection{Large-$\tau_s$ asymptotics}
\label{seclongtime}

For large $\tau$, Eq.~\eqref{eqlongcont} gives
\begin{equation}
	\bar G(\tau)\approx\sqrt{\dfrac{\tau}{2\pi}}.
	\label{eqlongmono}
\end{equation}
Substituting into Eq.~\eqref{eq:nondfinal_impact} gives
\begin{equation}
	Z''=\Lambda F, \qquad
	Y=-\frac{\Gamma}{\sqrt{2\pi}}
	\int_0^\tau F(\zeta)\sqrt{\tau-\zeta}\,{{\rm d}}\zeta,
	\qquad
	F=\alpha^{3/2}H(\alpha),
	\qquad
	\alpha=Y-Z.
	\label{eqlongred}
\end{equation}
We now introduce the further rescaling
\begin{equation}
	\tau=aT, \quad
	Y=b\hat Y, \quad
	Z=b\hat Z, \qquad \text{so that} \quad
	F=b^{3/2}\Phi,
	\quad
	\Phi=\hat\alpha^{3/2}H(\hat\alpha),
	\quad
	\hat\alpha=\hat Y-\hat Z,
	\label{eqlongscale}
\end{equation}
where $a,b$ are constants to be chosen, and primes on hatted
quantities within this section denote differentiation with respect to $T$.
Substituting Eq.\ \eqref{eqlongscale} into Eq.~\eqref{eqlongred} gives
\begin{equation}
	\hat Z''
	=
	\Lambda a^2b^{1/2}\Phi,
	\qquad
	\hat Y
	=
	-\frac{\Gamma}{\sqrt{2\pi}}
	a^{3/2}b^{1/2}
	\int_0^T
	\Phi(\zeta)\sqrt{T-\zeta}\,{\rm d}\zeta.
	\label{eqlongscaled}
\end{equation}
Choosing $a$ and $b$ so that the two prefactors equal unity gives
\begin{equation}
	a=\dfrac{\Gamma^2}{2\pi\Lambda^2},
	\qquad
	b=\dfrac{16\pi^4\Lambda^6}{\Gamma^8}.
	\label{eqlongab}
\end{equation}
The system then becomes the canonical, coefficient-free problem
\begin{equation}
	\hat Z''=\Phi, \qquad
	\hat Y=-\int_0^T
	\Phi(\zeta)\sqrt{T-\zeta}\,{\rm d}\zeta,
	\qquad
	\Phi=\hat\alpha^{3/2}H(\hat\alpha),
	\qquad
	\hat\alpha=\hat Y-\hat Z.
	\label{eqlongcanon}
\end{equation}
Since $Z'(0)=(b/a)\hat Z'(0)=-1$, the rescaled initial condition is
\begin{equation}
	\hat Z'(0)=-S_{\rm eff},
	\label{junk1}
\end{equation}
where
\begin{equation}
	S_{\rm eff}=\frac{a}{b}
	=\dfrac{\Gamma^{10}}{(2\pi)^5\Lambda^8}.
	\label{eqlongS}
\end{equation}
Thus, in this regime, $\Gamma$ and $\Lambda$ enter the solution only through the single parameter $S_{\rm eff}$. We note that this $S_{\rm eff}$ corresponds to the nondimensional parameter $S$ of \cite{BHATTACHARJEE2018147}, which is expected because the long-time impulse response of the Rayleigh beam matches that of the Euler Bernoulli beam to leading order. Note that Eqs.~\eqref{eqlongcanon} and \eqref{junk1} govern this scaled model of the impact, wherein the solution depends only on $S_{\rm eff}$. It follows that there is a critical $S_{\rm eff} = S_c$ such that (i) there is exactly one separation, and (ii) the impactor velocity at the separation is exactly zero. We cannot find $S_c$ analytically, but could find it numerically if we wished. By Eq.~\eqref{eqlongS}, then, we have
\begin{equation}
	\Gamma_c(\Lambda)
	=
	\left[(2\pi)^5S_c\right]^{1/10}\Lambda^{4/5}.
	\label{eqlongbnd}
\end{equation}
Note that the above case refers to impacts where there is a single and final separation, {\em and} the contact duration is long.

Furthermore, there are also $S_{\rm eff}$ values for which there is eventual sustained contact. For such cases, assuming $\tau_s$ to be large is always justifiable.

\begin{figure}[htbp]
	\centering
	\includegraphics[width=\textwidth]{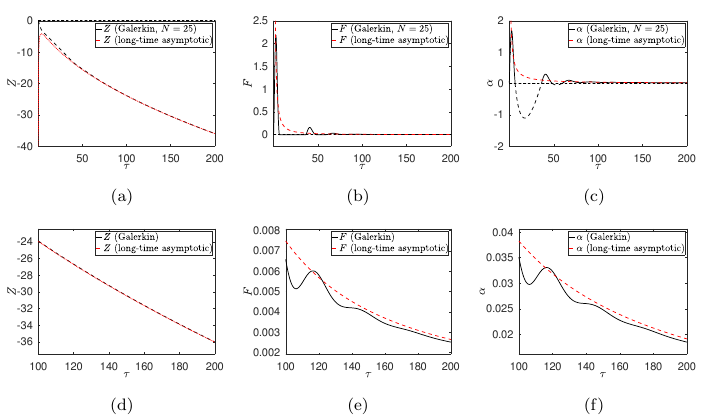}
	\caption{Comparison of the Galerkin solution with the long-time asymptotic expressions for $\Gamma=0.75$ and $\Lambda=0.1$ over $0\leq\tau\leq200$; (a) impactor displacement $Z$, (b) contact force $F$, and (c) indentation $\alpha$. Corresponding zoomed views are in (d), (e), and (f), respectively.}
	\label{fig:longtau}
\end{figure}

In this regime, regardless of how many separations preceded the final contact phase, we can focus on the final dynamics wherein $\hat\alpha\geq0$ is established. In this regime, also, because our two nondimensional parameters have collapsed into one, the asymptotic analysis of \cite{BHATTACHARJEE2018147} is directly applicable. Taking the final result from there
(their Eqs.~(36) and (47)), we have 
\begin{equation}
	\begin{aligned}
		&\hat Z(T)=-S_{\rm eff}\sqrt T+\frac{S_{\rm eff}\pi}{4}-\frac{S_{\rm eff}\pi}{8\sqrt T}+\cdots,\\
		&\hat\alpha(T)\sim\left(\frac{S_{\rm eff}}{4}\right)^{2/3}\frac1T.
	\end{aligned}
	\label{eqraycanonZ}
\end{equation}
Returning to our original nondimensional variables, we obtain
\begin{equation}
	\begin{aligned}
		&Z(\tau)=
		-\frac{\Gamma}{\Lambda}\sqrt{\frac{\tau}{2\pi}}
		+\frac{\Gamma^2}{8\Lambda^2}
		-\frac{\Gamma^3}{16\Lambda^3\sqrt{2\pi\tau}}+\cdots,\\
		&\alpha(\tau)\sim
		\left(\frac{\Gamma^2}{32\pi\Lambda^4}\right)^{1/3}\frac1\tau.
	\end{aligned}
	\label{eqrayzphys}
\end{equation}
The contact force follows from $F=\alpha^{3/2}$ as
\begin{equation}
	F(\tau)\sim
	\left(\frac{\Gamma^2}{32\pi\Lambda^4}\right)^{1/2}\frac1{\tau^{3/2}},
	\label{eqrayFphys}
\end{equation}
which goes to zero as $\tau\to\infty$.
Interestingly, the above three results are predicted to be independent of the number and nature of intervening contact events. The prediction is borne out in Fig.~\ref{fig:longtau}, for $\Gamma=0.75$ and $\Lambda=0.1$. $Z$, $F$, and $\alpha$ indeed approach their predicted asymptotic behaviours for large $\tau$, although there are complex preceding contact and separation events.

\subsection{Short-$\tau_s$ asymptotics}
\label{secshorttime}

For small $\tau$, Eq.~\eqref{eqshortcont} gives
\begin{equation}
	\bar G(\tau)\approx\dfrac{\tau}{2}.
	\label{eqshortmono}
\end{equation}
Substituting into Eq.~\eqref{eq:nondfinal_impact} gives
\begin{equation}
	Z''=\Lambda F,\qquad
	Y=-\frac{\Gamma}{2}\int_0^\tau F(\zeta)(\tau-\zeta)\,{\rm d}\zeta,
	\qquad
	F=\alpha^{3/2}H(\alpha),\qquad
	\alpha=Y-Z.
	\label{eqshortred}
\end{equation}
Again introducing further rescaling,
\begin{equation}
	\tau=aT, \quad
	Y=b\hat Y, \quad
	Z=b\hat Z,  \qquad \text{so that} \quad
	F=b^{3/2}\Phi,
	\quad
	\Phi=\hat\alpha^{3/2}H(\hat\alpha),
	\quad
	\hat\alpha=\hat Y-\hat Z.
	\label{eqshortscale}
\end{equation}
Substituting Eq.\ \eqref{eqshortscale} into Eq.\ \eqref{eqshortred} gives
\begin{equation}
	\hat Z''
	=
	\Lambda a^2b^{1/2}\Phi,
	\qquad
	\hat Y
	=
	-\frac{\Gamma}{2}a^2b^{1/2}
	\int_0^T\Phi(\zeta)(T-\zeta)\,{\rm d}\zeta.
	\label{eqshortscaled}
\end{equation}
Both equations contain the same factor $a^2b^{1/2}$, wherein we have two free rescaling parameters. We specify $\Lambda a^2b^{1/2}=1$ and use the remaining free parameter to set the rescaled initial velocity
to $\hat Z'(0)=-1$. We easily obtain
\begin{equation}
	a=b=\Lambda^{-2/5}.
	\label{eqshortsys}
\end{equation}
Also, $\hat Z''=\Phi=\hat\alpha^{3/2}$ during the contact. Our system is therefore
\begin{equation}
	\hat Z''=\Phi, \quad
	\hat Y=-R\int_0^T\Phi(\zeta)(T-\zeta)\,{\rm d}\zeta,
	\quad
	\Phi=\hat\alpha^{3/2},
	\quad
	\hat\alpha=\hat Y-\hat Z,
	\quad
	\hat Z'(0)=-1 \quad \hat Y(0)=\hat Z(0)=0,
	\label{eqshortcanon}
\end{equation}
where the two original nondimensional parameters have collapsed into one:
\begin{equation}
	R=\dfrac{\Gamma}{2\Lambda}.
	\label{eqshortR}
\end{equation}

\begin{figure}[htbp]
	\centering
	\includegraphics[width=0.4\textwidth]{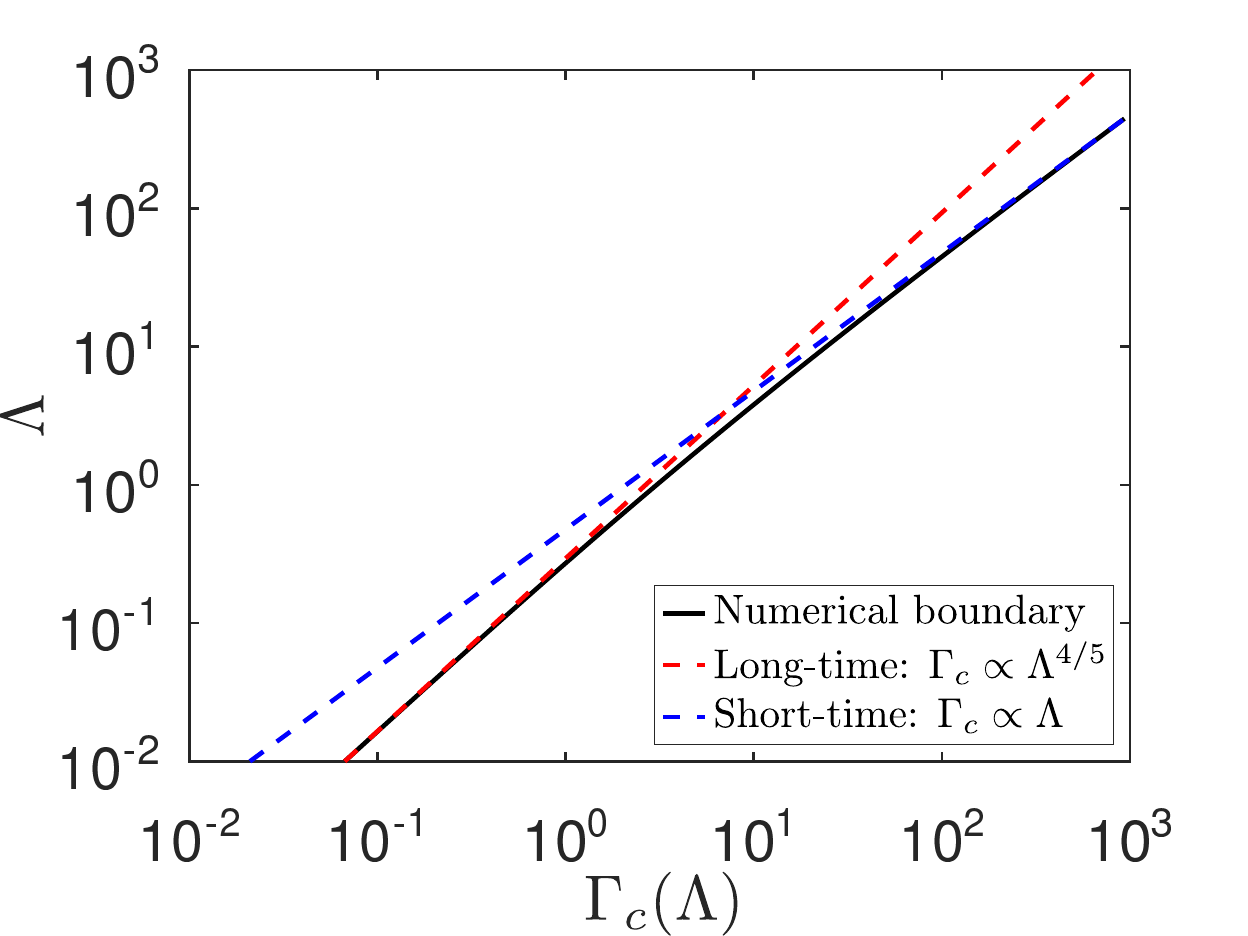}
	\caption{Numerically estimated $\Gamma_c(\Lambda)$. The dashed lines have
		the two asymptotic slopes.}
	\label{fig:slopecomb}
\end{figure}

Now we note that, for small $\tau_s$ to determine the final dynamics, the impact must necessarily end with a quick first separation. Presumably, even among such separations, we can seek those that end with the impactor velocity being exactly zero. Proceeding on the assumption that such impact terminations exist, we can in principle find the corresponding value of $R$, say $R_c$, through numerical solution and iterative search.
We immediately obtain, from Eq.~\eqref{eqshortR},
\begin{equation}
	\Gamma_c = 2 R_c \Lambda.
	\label{junk2}
\end{equation}

We thus have two limiting regimes for the critical parameter $\Gamma_c$, which is the boundary between separation and eventual sustained contact. For impacts that end in quick separation, we have Eq.~\eqref{junk2}. For eventual sustained impact, we have Eq.~\eqref{eqlongbnd}. These two predicted power law behaviours in terms of $\Lambda$ are consistent with numerics: see Figure \ref{fig:slopecomb}.


\section{Concluding remarks}
\label{conc}

We have studied the dynamics of a weightless Hertzian impactor striking an infinitely long Rayleigh beam. Unlike the Timoshenko beam, the Rayleigh beam has an impulse response wherein the struck point's velocity is bounded just after impact. The impact problem is formulated using point contact between the beam and the impactor, with contact mediated by a Hertzian ($3/2$-power) spring. The system is governed by two nondimensional parameters, but in the two limits of very short and very long contact these two parameters collapse into different single nondimensional parameters.

Beyond the academic interest of the studied impact problem itself, the results obtained may apply well to impacts with finitely long beams because wave speeds in the Rayleigh beam are bounded and the impact interaction can be over before reflections from the boundary have time to return.

A significant new contribution of this paper is in recasting the impulse response as an integral using an abstract parameter (not equivalent to wave number, as in the formula obtainable using frequency domain methods) which allows Galerkin approximation with relatively fewer states in a local-in-time ODE model. After establishing the numerical accuracy of this approach to discretizing the dynamics, we have been able to present a complete qualitative picture of the possible modes of impact interaction. There are three distinct regimes in the impact behaviour. We may observe a single contact phase followed by rebound (no further contact), or a single sustained contact phase that continues for all time, or more than one periods of intermediate separation followed by sustained contact. Some simple asymptotic descriptions of relevant variables are obtainable in limiting cases, and are found to be consistent with numerics.

Over and above the insights provided into the dynamics of this particular problem, we hope that our indirect Galerkin formulation may find interest in the research community for use in other problems involving convolution integrals.


\bibliographystyle{unsrt}
\bibliography{cas-refs}

\end{document}